\documentclass[aps,prl,twocolumn,groupedaddress,draft,showpacs]{revtex4}
\usepackage{amsmath}

\newcommand{\beq}{\begin{equation}}
\newcommand{\eeq}{\end{equation}}
  \newcommand{\beql}[1]{\begin{equation}\label{eq:#1}}
  \newcommand{\beqa}{\begin{eqnarray}}
  \newcommand{\eeqa}{\end{eqnarray}}
  \newcommand{\beqas}{\begin{eqnarray*}}
  \newcommand{\eeqas}{\end{eqnarray*}}

  \newcommand*{\R}{\mathbf{R}}
  
 \newcommand*{\dd}{\mathrm{d}}
 \newcommand*{\bP}{\mathbf{P}}
  \newcommand*{\bS}{\mathbf{S}}
  \newcommand*{\cF}{\mathcal{F}}
  \newcommand*{\cH}{\mathcal{H}}
  \newcommand*{\cK}{\mathcal{K}}
  \newcommand*{\cP}{\mathcal{P}}
  \newcommand*{\al}{\alpha}
  \newcommand*{\be}{\beta} 
  \newcommand*{\ep}{\epsilon}
   \newcommand*{\eq}[1]{(\ref{eq:#1})}
  \newcommand*{\et}{\eta}
  \newcommand*{\ga}{\gamma}
  \newcommand*{\nn}{\nonumber}
  \newcommand*{\om}{\omega}
  \newcommand*{\ps}{\psi} 
 \newcommand*{\rh}{\rho}
  \newcommand*{\si}{\sigma} 
  \newcommand*{\ve}{\varepsilon}
  \newcommand*{\De}{\Delta}                                          
  \newcommand*{\Eq}[1]{Eq.~(\ref{eq:#1})}
  \newcommand*{\Om}{\Omega}
  \newcommand*{\Th}{\Theta}

\newcommand*{\bracket}[1]{\langle#1\rangle}

\newcommand{\oep}{\overline{\varepsilon}}
\newcommand{\oet}{\overline{\eta}}
\newcommand{\uep}{\overline{\ve}}
\newcommand{\uet}{\overline{\et}}
\newcommand{\x}{Q}
\newcommand{\px}{P}
\newcommand{\y}{\overline{Q}}
\newcommand{\py}{\overline{P}}
\newcommand{\bq}{q}
\newcommand{\bp}{p}
\renewcommand{\o}{\om}

\begin{document}
\title{Disproving Heisenberg's error-disturbance relation}
\author{Masanao Ozawa}
\affiliation{Graduate School of Information Science,
Nagoya University, Chikusa-ku, Nagoya, 464-8601, Japan}
\begin{abstract}
Recently, Busch, Lahti, and Werner \cite{BLW13} claimed that Heisenberg's 
error-disturbance relation can be proved  in its original form with new 
formulations of error and disturbance, 
in contrast to the theory proposed by the present author 
\cite{{02KB5E},{03HUR},{03UVR},{04URN}} 
and confirmed by recent experiments \cite{12EDU,RDMHSS12,13EVR,13VHE}.
Despite their claim, it is shown here that a class of solvable models of 
position measurement with explicit interaction Hamiltonians escape 
the Busch-Lahti-Werner relation.
It is also made clear where their proof fails.
Those models have unambiguously defined zero root-mean-square error 
and finite root-mean-square disturbance in every input state
and are naturally considered to violate Heisenberg's error-disturbance relation 
in any conceivable formulation.
\end{abstract}
\maketitle

In his seminal paper \cite{Hei27}  Heisenberg in 1927 introduced 
his error-disturbance relation (EDR) through the famous $\gamma$ ray microscope 
thought experiment.
Let $\x$ be a coordinate of a particle to be measured and $\px$ its momentum.
He considered the ``mean error'' $\ve(\x)$ with which the observable $\x$ is measured 
and the ``mean disturbance'' $\et(\px)$ with which the observable $\px$ is changed by
the measuring interaction.
Under some implicit assumptions about the post-measurement state
\footnote{As one of the simplest interpretations,  those hidden assumptions are
given as the conjunction of conditions (i) and (ii) as follows.
(i) If $\x$ and $\px$ are simultaneously measured with mean errors $\ve(\x)$ and
$\ve(\px)$, the post-measurement standard deviations $\si(\x)$ and $\si(\px)$ satisfy
$$
\ve(\x)\ge\si(\x),\quad\mbox{and}\quad \ve(\px)\ge\si(\px). 
$$
(ii) If $\x$ is measured with mean error $\ve(\x)$ and mean momentum disturbance 
$\et(\px)$,  then $\x$ and $\px$ can be simultaneously measured with mean errors $\ve(\x)$ and
$\et(\px)$.
Here, the standard deviation $\si(A)$ is defined  by 
$\si(A)=\bracket{(A-\bracket{A})^2}^{1/2}$ 
for an arbitrary observable $A$,
where $\bracket{\cdots}$ stands for the mean in a given state. 
For more discussions see Ref.~\cite{03HUR}.
},
he derived his EDR 
\beql{HUP}
\ve(\x)\et(\px)\ge \frac{\hbar}{2}
\eeq
from Kennard's relation
\beql{Ken27}
\si(\x)\si(\px)\ge\frac{\hbar}{2}
\eeq
for the standard deviations $\si(\x),\si(\px)$ of $\x,\px$. 
Note that Heisenberg \cite{Hei27} proved \Eq{Ken27} 
for Gaussian wave functions, and
Kennard \cite{Ken27} gave a complete proof subsequently.

In 1980, Braginsky, Vorontsov, and Thorne \cite{BVT80} claimed that Heisenberg's
EDR leads to a sensitivity limit, 
called the standard quantum limit, for gravitational wave detectors. 
However, following Yuen's \cite{Yue83} proposal of exploiting ``contractive states,'' 
the present author \cite{88MS,89RS} in 1988 
constructed a solvable model of ``error-free contractive-state (position) measurement'' 
that breaks the standard quantum limit
and violates Heisenberg's EDR in any state \cite{02KB5E}.
Now,  Heisenberg's EDR and its consequences 
are taken to be breakable limits
\cite{02KB5E,03HUR,03UVR,04URN,GLM04,12EDU,RDMHSS12,Bra13,13EVR,13VHE}.

In contrast, Busch, Lahti, and Werner (BLW) \cite{BLW13} recently claimed 
that the failure of Heisenberg's EDR is due to a wrong definition of momentum disturbance,
and that Heisenberg's EDR can be generally proved with a new formulation.
Based on an exhaustive study of error and disturbance of linear (position) measurements  \cite{90QP}
including the error-free contractive-state measurement \cite{88MS}, 
in this note it will be shown that BLW's criticism on the definition of momentum 
disturbance is groundless and that their proof of Heisenberg's EDR fails.

Heisenberg's original formulation \cite{Hei27} explicitly concerns the ``mean error'' and 
the ``mean disturbance.'' 
In our approach they are rigorously defined as the 
``root-mean-square error'' and the ``root-mean-square disturbance'' 
\cite{02KB5E,03HUR,03UVR,04URN}. 
BLW criticized our definitions claiming that the momentum before and after the 
measuring interaction do not commute, so that difference makes no sense \cite{BLW13}.  
However, as we will see, BLW neglected the fact that for linear measurements,
the momentum before and after the measuring interaction, as well as the pre-measurement 
position and the post-measurement meter position, do commute.  
In that case, the ``root-mean-square error'' and the ``root-mean-square disturbance'' 
are unambiguously determined from classical definitions.
Actually all the error-free linear measurements 
including the error-free contractive-state measurement have zero root-mean-square error 
and finite root-mean-square disturbance according to the classical definitions.
Thus, the failure of Heisenberg's relation is not due to a wrong definition 
of momentum disturbance.

In order to retain the original form of EDR, BLW~\cite{BLW13} redefined 
the EDR as the relation between the supremums of the ``mean error'' 
and the ``mean disturbance'' over a large class of input states and claimed that 
Heisenberg's EDR can be proved in its original form
with their new formulation.
With mathematically careful examinations it is shown that their alleged proof includes a 
loophole.  In fact, all the error-free linear measurements do not satisfy BLW's 
relation \cite{BLW13}.

It will also be shown that BLW's error and disturbance are infinite for almost all linear 
measurements, and by no means captures the physical content of Heisenberg's original 
discussion.
In addition to the recent criticisms \cite{RMHS13} on their formulation 
from an interpretational point of view, those results suggest
that the claim made by Busch, Lahti, and Werner \cite{BLW13} is unsupported.

\paragraph{Linear measurements.}
Consider a one-dimensional mass, called an {\em object},
with position $\x$ and momentum $\px$, described by a Hilbert space $\cH$.
A measurement of $\x$ using 
a {\em probe} $\bP$ is described as follows.  The probe $\bP$ is another one-dimensional mass
with position $\y$ and momentum $\py$, described by a Hilbert space $\cK$.
The measurement of $\x$ is carried out by coupling between $\bS$ and $\bP$
turned on from time $t=0$ to $t=\De t$,  and the outcome of the measurement 
is obtained by measuring the probe position $\y$, called the {\em
meter observable}, at time $t=\De t$.
The total Hamiltonian for the object and the probe is taken to be
                                                                \beql{(1)}
{H}_{\bS+\bP} = {H}_{\bS} + {H}_{\bP} + K{H},
                                                                \eeq
where ${H}_{\bS}$ and ${H}_{\bP}$ are the free Hamiltonians
of $\bS$ and $\bP$, respectively, ${H}$ represents the measuring interaction,
and $K$ is the coupling constant.  We assume that the coupling
is so strong 
that ${H}_{\bS}$ and ${H}_{\bP}$ can be neglected.
We choose $\De t$ as $K\De t=1$. 
We suppose that, possibly by the linear approximation, the measuring 
interaction $H=H(\al,\be,\ga)$ is given by
                                                                \beql{(2)}
{H}(\al,\be,\ga) = \al (\x\px-\y{\py})
                             +\be \y\px+\ga \x{\py},
                                                                \eeq
where $\al ,\ \be ,\ \ga  \in \R$ \cite{90QP}. 
In this case, the measurement  will be referred to as a {\em linear (position) measurement}.

The model with $(\al,\be,\ga)=(0,0,1)$ has been know as von Neumann's model of
position measurement \cite{vN55}.
The model with $(\al,\be,\ga)=(1,-2,2)/(3\sqrt{3})$ has been know as the error-free
contractive state measurement \cite{88MS,89RS}.

Solving Heisenberg's equations of motion,
we have \cite{90QP}
\beqa
\x(\De t)&=&a\x(0)+b \y(0),\label{eq:S1}\\
\y(\De t)&=&c \x(0)+d \y(0),\label{eq:S2}\\
\px(\De t)&=&d\px(0)-c \py(0),\label{eq:S3}\\
\py(\De t)&=&-b\px(0)+a \py(0),\label{eq:S4}
\eeqa                         
where  $a,b,c,d$ satisfy
\beql{det}
ad - bc = 1,
\eeq 
and are given as follows.

(i) If $\al^2+\be\ga=0$, we have
                                                \beql{sol-1}
\left[\begin{array}{cc} a & b   \\       
                                 c & d \end{array}\right]
=
                  \left[\begin{array}{cc} \al& \be   \\       
                                 \ga & -\al \end{array}\right]
+
\left[\begin{array}{cc} 1& 0   \\       
                                   0 &  1 \end{array}\right].               
                                 \eeq                                          
                                 
(ii) If $\al^2+\be\ga<0$, letting $D=\sqrt{-(\al^2+\be\ga)}$ we have
\beql{sol-2}
\left[\begin{array}{cc} a & b   \\       
                                 c & d \end{array}\right]
=
                                 \frac{\sin D}{D}
                                 \left[\begin{array}{cc} \al& \be   \\       
                                 \ga & -\al \end{array}\right]
+\cos D
\left[\begin{array}{cc} 1& 0   \\       
                                   0 &  1 \end{array}\right].                      
\eeq

(iii) If $\al^2+\be\ga>0$, letting $E=\sqrt{\al^2+\be\ga}$ we have
\beql{sol-3}
\left[\begin{array}{cc} a & b   \\       
                                 c & d \end{array}\right]
=
                        \frac{\sinh E}{E}
                                 \left[\begin{array}{cc} \al& \be   \\       
                                 \ga & -\al \end{array}\right]
                                 +
\cosh E
\left[\begin{array}{cc} 1& 0   \\       
                                   0 &  1 \end{array}\right].
\eeq

Let
$U(\al,\be,\ga)=\exp[-iH(\al,\be,\ga)/\hbar]$.
We denote the above model by $(\cK,\xi,U(\al,\be,\ga),\y)$, 
where $\xi$ stands for
the state of the probe at $t=0$ \footnote{
See Ref.~\cite{84QC} for general theory of measuring processes.
See Ref.~\cite{89RS} for general theory of position measurements.
}.

In order to avoid using physically inaccessible resource to break Heisenberg's error-disturbance
relation, we assume that the state $\xi$ of $\bP$ at $t=0$ satisfies the condition that
the wave function $\xi(q)$ is infinitely differentiable and 
$\|\y^{n}\py^{m}\xi\|<\infty$ for all $m,n$.
We also assume that the state $\ps$ of $\bS$ at $t=0$ satisfies the analogous condition 
\footnote{
Mathematically, the above conditions are equivalent to that the wave functions 
$\xi(q)$ and $\psi(q)$ are Schwartz rapidly decreasing functions \cite{RS80}.
}.

\paragraph{Root-Mean-Square Error and Disturbance.}
In order to define ``mean error'' and ``mean disturbance'' for this measurement,
we recall classical definitions.
Suppose that the true value is given by $\Theta=\theta$ and 
its measured value is given by $\Om=\om$.   
For each pair of values $(\Th,\Om)=(\theta,\om)$, 
the error is defined as $\theta-\om$.  
To define the ``mean error'' with respect to the joint probability distribution 
$ \mu^{\Th,\Om}(d\theta,d\om)$ of $\Th$ and $\Om$, 
Gauss \cite{Gau95} introduced the {\em root-mean-square error} 
$\ve_{G}(\Th,\Om)$ of $\Om$ for $\Th$ as
\beql{rmse}
\ve_{G}(\Th,\Om)=\left(\iint_{\R^{2}}(\om-\theta)^{2} \mu^{\Th,\Om}(\dd\theta,\dd\om)\right)^{1/2},
\eeq
which Gauss  \cite{Gau95} called the ``mean error'' or the ``mean error to be feared'' \footnote{
Gauss actually assumed the probability distribution of $\Th-\Om$, which is naturally derived
from the joint probability distribution of $\Th$ and $\Om$.
},
and has long been accepted as a standard definition for  the ``mean error.'' 

In the model  $(\cK,\xi,U(\al,\be,\ga),\y)$, the value of the observable $\x(0)$ at $t=0$ is 
measured by the value of the meter observable $\y(\De t)$ at $t=\De t$.  
Since $\x(0)$ and $\y(\De t)$ commute, as seen from \Eq{S2},
we have the joint 
probability distribution $\mu^{\x(0),\y(\De t)}(\dd\theta,\dd\o)$ of the values of $\x(0)$ and $\y(\De t)$
as
\beq
\mu^{\x(0),\y(\De t)}(\dd\theta,\dd\o)=\bracket{E^{\x(0)}(\dd\theta)E^{\y(\De t)}(\dd\o)},
\eeq
where $E^{A}$ stands for the spectral measure of an observable $A$ \cite{Hal51}, 
and  $\bracket{\cdots}$ stands for the mean value in the state $\ps\otimes\xi$.
Then,  from \Eq{rmse} the {\em root-mean-square error} $\ve(\x,\ps)$
of $\y(\De t)$  for  $\x(0)$ in $\ps$ is given by 
\beqa
\ve(\x,\ps)&=&\ve_G(\x(0),\y(\De t))\nn\\
&=&\left(\iint_{\R^{2}}(\o-\theta)^{2}\mu^{\x(0),\y(\De t)}(\dd\theta,\dd\o)\right)^{1/2}\nn\\
&=&\bracket{(\y(\De t)-\x(0))^{2}}^{1/2}\\
&=&\| [(c-1)\x(0)+d \y(0)](\psi\otimes\xi)\|<\infty.
\label{eq:E-V}
\eeqa

Since $\px(0)$ and $\px(\De t)$ commute, as seen from \Eq{S3},
we have the joint 
probability distribution $\mu^{\px(0),\px(\De t)}(\dd\theta,\dd\o)$ of the values of $\px(0)$ and $\px(\De t)$
as
\beq
\mu^{\px(0),\px(\De t)}(\dd\theta,\dd\o)=\bracket{E^{\px(0)}(\dd\theta)E^{\px(\De t)}(\dd\o)}.
\eeq
Then,  the {\em root-mean-square disturbance} $\et(\px,\ps)$ of $\px$ from $t=0$ to
$t=\De t$ is defined as the root-mean-square error of $\px(\De t)$ for $\px(0)$ given by 
\beqa
\et(\px,\ps)&=&\ve_G(\px(0),\px(\De t))\nn\\
&=&\left(\iint_{\R^{2}}(\o-\theta)^{2}\mu^{\px(0),\px(\De t)}(\dd\theta,\dd\o)\right)^{1/2}\nn\\
&=&\bracket{(\px(\De t)-\px(0))^{2}}^{1/2}\\
&=&
\| [(d-1)\px(0)-c \py(0)](\psi\otimes\xi)\|<\infty.
\label{eq:D-V}
\eeqa

\paragraph{Heisenberg's EDR.}
We have
\beqa
\lefteqn{
[(c-1)\x(0)+d\y(0),(d-1)\px(0)-c\py(0)]}\qquad\qquad\qquad\qquad\qquad
\qquad
\nn\\
&=&(1-c-d)i\hbar.
\eeqa
By Eqs.~\eq{E-V}, \eq{D-V}, and the Schwarz inequality, we have
\beqa
\ve(\x,\psi)\et(\px,\psi)\ge\frac{|1-c-d|\hbar}{2}.
\eeqa
Consequently, 
if $c+d\le 0$ or $2\le c+d$,  we have
\beqa
\ve(\x,\psi)\et(\px,\psi)\ge\frac{\hbar}{2},
\eeqa
so that Heisenberg's EDR holds for 
every $\ps$.

\paragraph{Error-free linear measurements.}
From \Eq{E-V}, 
if 
$c=1$
and 
$d=0$,
we have
\beq
\ve(\x,\psi)=0
\eeq
for all $\ps$.
From \Eq{det},  the constraint on $a,b,c,d$ is given by
\beqa\label{eq:constraint}
b=-1, \quad c=1, \quad d=0, \quad a=\mbox{arbitrary,}
\eeqa
and hence we have
\beqa
\x(\De t)&=&a\x(0)-\y(0),\\
\y(\De t)&=&\x(0),\\
\px(\De t)&=&-\py(0),\\
\py(\De t)&=&\px(0)+a\py(0).
\eeqa
In this case, we have
\beqa
\et(\px,\psi)^{2}&\!=\!&\| (\px(0)+\py(0))(\psi\otimes\xi)\|^{2}\nn\\
&\!=\!&\si(\px)^2+\si(\py)^2+(\bracket{\ps|\px|\ps}+\bracket{\xi|\py|\xi})^{2},\nn\\
\label{eq:etp}
\eeqa
so that the disturbance is independent of $a$.

From Eqs.~\eq{sol-1}--\eq{sol-3},
 the Hamiltonian $H=H(a)$ realizing this model is given by
\beql{H-a}
H(a)=\Om(a)\left\{\frac{a}{2}(\x \px-\y\py)-\y\px+\x\py\right\}
\eeq
for any $a>-2$,
where
\beql{O-a}
\Om(a)=\left\{
\begin{array}{cl}
\displaystyle\frac{\cos^{-1}\frac{a}{2}}{\sqrt{1-(\frac{a}{2})^2}} &(-2<a<2),\\
1&({a}=2),\\
\displaystyle\frac{\cosh^{-1}\frac{a}{2}}{\sqrt{(\frac{a}{2})^2-1}} &(2<a).
\end{array}\right.
\eeq

Let
$
U(a)=\exp[-iH(a)/\hbar].
$
The model $(\cK,\xi,U(a),\y)$ is called the {\em error-free linear 
measurement} for $a>-2$.  The error-free contractive-state 
measurement corresponds to the case where $a=1$. 

Since $\ve(\x,\ps)=0$ and $\et(\px,\ps)<\infty$,
we have 
\beq
\ve(\x,\ps)\et(\px,\ps)=0
\eeq
for all $\psi$.
Thus, every error-free linear measurement violates
Heisenberg's EDR in any input state.

\paragraph{Appleby's formulation.}
To retain the original form of Heisenberg's EDR, 
Appleby \cite{App98c} in 1998 considered their supremums over all input states, 
instead of the ``mean error'' and the ``mean disturbance'' in any state.
He defined the {\em uniform rms error} $\oep(\x)$ and the {\em uniform rms 
disturbance} $\oet(\px)$ by
\beqas
\oep(\x)&=&\sup_{\ps}\ve(\x,\ps),\\
\oet(\px)&=&\sup_{\ps}\et(\px,\ps),
\eeqas
where the supremum is taken over all input states $\psi$.
Then, Appleby \cite{App98c} sketched a proof of the relation
\beql{AHUP}
\oep(\x)\oet(\px)\ge\frac{\hbar}{2}.
\eeq

This relation is not universally valid even for linear measurements.
We have already seen that all the models $(\cK,\xi,U(a),\y)$ with $a>-2$ satisfy
$\oep(\x)=0$.
From \Eq{etp} we have $\oet(\px)=\infty$.
Thus, the product $\oep(\x)\oet(\px)$ is
indeterminate and cannot be concluded to be above $\hbar/2$.
Therefore,  Appleby's formulation of Heisenberg's EDR, \Eq{AHUP}, does not hold 
for models $(\cK,\xi,U(a),\y)$ with $a>-2$.

In fact, the indeterminate product $\oep(\x)\oet(\px)$ can be practically observed to 
be zero,  since we can practically check only a finite number 
of different states, and we have
\beq
\sup_{F\in\cF}[\sup_{\ps\in F}\ve(\x,\ps)\sup_{\ps\in F}\et(\px,\ps)]=0,
\eeq
where $\cF$ is the totality of finite sets $F$ of states $\ps$.

Another criticism against this approach is that we almost always have
$\uep(\x)=\uet(\px)=+\infty$.
More precisely, this happens for all $(\cK,\xi,U(\al,\be,\ga),\y)$ 
except for the set of parameters $(\al,\be,\ga)$ of Legesgue measure zero.
In fact, from \Eq{E-V}, we have
\beql{UEPQ}
\uep(\x)=\left\{
\begin{array}{ll}
d\| \y\xi\|& (c=1),\\
+\infty& (c\ne 1).
\end{array}
\right.
\eeq
From \Eq{D-V} we also have
\beql{UEPQ}
\uet(\px)=\left\{
\begin{array}{ll}
c\| \py\xi\|& (d=1),\\
+\infty& (d\ne 1).
\end{array}
\right.
\eeq

\paragraph{BLW's formulation.}
Now we turn to BLW's proposal \cite{BLW13}.
Here, we shall show that BLW's formulation is equivalent to Appleby's formulation
for any linear measurements, and hence we conclude that BLW's
formulation of Heisenberg's EDR
does not hold for models $(\cK,\xi,U(a),\y)$ for all $a>-2$.
Moreover, BLW's position-measurement-error and momentum-disturbance 
are almost always infinite.

For any model $(\cK,\xi,U(\al,\be,\ga),\y)$ the POVM $\Pi^{\x'}(\dd\o)$ is defined by
$$
\Pi^{\x'}(\dd\o)=\bracket{\xi|E^{\y(\De t)}(\dd\o)|\xi}.
$$
For any state $\psi$ and any real number $\theta$
the root-mean-square deviation of $\Pi^{\x'}$ from $\theta$ in $\psi$ is defined by
$$
D(\psi,\x';\theta)^2=\int_{\R} (\o-\theta)^2 \bracket{\psi|\Pi^{\x'}(\dd\o)|\psi}.
$$
Then, BLW's error $\De_{c}(\x,\x')$ of $\Pi^{\x'}$ for $\x$ is defined by \footnote{
BLW \cite{BLW13} originally defined $\De_{c}(\x,\x';\ep)$ as the supremum
of $D(\rh,\x';\theta)$ similarly defined for density operators $\rh$, but that is equivalent 
to the present formulation by the convexity of the function 
$\rh\mapsto D(\rh,\x';\theta)^2$.}
\begin{eqnarray*}
\De_{c}(\x,\x')
&=&\lim_{\ep\to 0}\De_{c}(\x,\x';\ep),\\
\De_{c}(\x,\x';\ep)&=&\sup\{D(\ps,\x';\theta)|\psi, \theta; \|\x\psi-\theta\psi\|\le \ep\}.
\end{eqnarray*}
With the POVM $\Pi^{\px'}$ defined by
$$
\Pi^{\px'}(\dd\o)=\bracket{\xi|E^{B(\De t)}(\dd\o)|\xi},
$$
BLW's disturbance $\De_{c}(\px,\px')$ of $\px$ for the measurement $(\cK,\xi,U(\al,\be,\ga),\y)$
is defined analogously.
Then, BLW \cite{BLW13}
claimed that the relation
\beql{BLW}
\De_{c}(\x,\x')\De_{c}(\px,\px')\ge\frac{\hbar}{2}
\eeq
can be proved generally.

However, we can prove that for any linear measurement 
$(\cK,\xi,U(\al,\be,\ga),\y)$ 
we have
 \beqa
 \De_{c}(\x,\x')&=&\uep(\x),\label{eq:BLWQ}\\
 \De_{c}(\px,\px')&=&\uet(\px).\label{eq:BLWP}
 \eeqa
Thus, BLW's formulation is equivalent to Appleby's formulation, 
and we conclude that \Eq{BLW} does not hold for all the error-free
linear measurements $(\cK,\xi,H(a),\y)$ with $a>-2$.
Moreover, we have $ \De_{c}(\x,\x')= \De_{c}(\px,\px')=+\infty$
for all $(\cK,\xi,U(\al,\be,\ga),\y)$ except for the set of parameters 
$(\al,\be,\ga)$ of Lebesgue measure zero.

The proof runs as follows.  We have
\beqa
D(\ps,\x';\theta)&=&\|\y(\De t)\ps\otimes\xi-\theta\ps\otimes\xi\|,\\
\ve(\x,\psi)&=&\|\y(\De t)\ps\otimes\xi-\x(0)\ps\otimes\xi\|,
\eeqa
so that by the triangular inequality, if $\|\x\psi-\theta\psi\|\le \ep$, we have
\beql{DEP}
|D(\ps,\x';\theta)-\ve(\x,\psi)|\le \ep.
\eeq
Let
\beqas
D(\x,\x';\ep)&=&\sup\{\ve(\x,\psi)|\ps,\theta; \|\x\psi-\theta\psi\|\le \ep\}.
\eeqas
We have
\beql{DCD}
|\De_{c}(\x,\x';\ep)-D(\x,\x';\ep)|\le \ep.
\eeq
It follows from \Eq{E-V} that $\sup_{\ps}\ve(\x,\ps)$ is determined 
by the values of $\ve(\x,\ps)$ only for approximate eigenstates $\ps$ of $Q$
\footnote{See Ref.\ \cite[p.\ 51]{Hal51} for approximate point spectrum.}, 
so that we have
\beql{DUEP}
D(\x,\x';\ep)=\uep(\x)
\eeq
for every $\ep>0$.
Thus, \Eq{BLWQ} follows from Eqs.~\eq{DCD} and \eq{DUEP}.
The proof of \Eq{BLWP} is similar.

\paragraph{Where does BLW's proof fail?}
As explained in Ref.~\cite{BLW13}, BLW's proof has two parts.
In the first part, it is shown that every covariant joint POVM satisfies
\Eq{BLW}.  In the second part, BLW claims that it can be shown that for any joint POVM
$\Pi$, there is a covariant one, say $\overline{\Pi}$, with marginals having at most 
the same $\De_{c}$'s.  However, BLW's proof has failed here, since no covariant POVM
$\overline{\Pi}$ satisfies $\De_{c}(\x,\Pi^{\x'})=0$.  The exact point where the proof
fails is examined as follows.

According to Ref.~\cite{BLW13}, BLW introduce the set
$\cP_{\ep}(\De\x,\De\px)$ of joint POVMs $\Pi$ such that
$D(\psi,\x';\theta)\le\De\x$ if $\|\x\psi-\theta\psi\|\le\ep$ and that 
$D(\psi,\px';\theta)\le\De\px$ if $\|\px\psi-\theta\psi\|\le\ep$.
They claimed that this is a compact convex set in a suitable weak topology. 
However, this is not true.  The compact convex closure $\overline{\cP}$ of 
$\cP_{\ep}(\De\x,\De\px)$ includes those joint POVMs on the
Stone-\v{C}ech compactification $\overline{\R^{2}}$ of $\R^2$
\footnote{Here, $\cP_{\ep}(\De\x,\De\px)$ is embedded in the space
of POVMs on $\overline{\R^{2}}$.}.
Thus, the Markov-Kakutani fixed theorem should 
be applied to $\overline{\cP}$ instead of $\cP_{\ep}(\De\x,\De\px)$.
Then,  there is a case where the covariant 
element $\overline{\Pi}$ in $\overline{\cP}$ is not a joint POVM on $\R^{2}$ and
it may happen that $\overline{\Pi}(\R^2)=0$.

This latter case indeed happens for the joint POVM $\Pi$ associated with error-free 
linear measurements.  In this case, the joint POVM $\Pi$ is uniquely defined 
for all $(\cK,\xi,U(a),\x)$ by
\beqa
\Pi(\dd\bq,\dd\bp)&=&\bracket{\xi|E^{\x(\De t)}(\dd\bq)E^{p(\De t)}(\dd\bp)|\xi}\nn\\
&=&\bracket{\xi|E^{\x}(\dd\bq)\otimes E^{-\px}(\dd\bp))|\xi}\nn\\
&=&|\hat{\xi}(-p)|^{2}E^{\x}(\dd\bq)\dd\bp,
\eeqa
where $\hat{\xi}$ is the Fourier transform of $\xi$.
According to
the Markov-Kakutani fixed point theorem, the fixed points $\overline{\Pi}$
actually exist in $\overline{\cP}$ but they satisfy
$\overline{\Pi}(\R^2)=0$.  Thus, there is no covariant joint POVM $\overline{\Pi}$
on $\R^{2}$ with marginals having at most the same $\De_{c}$'s as $\Pi$. 
In fact, for any fixed-point $\overline{\Pi}$ there is an invariant mean $m$ \cite{Gre69}
on 
the space $C(\R)$ of bounded continuous functions on $\R$ such that
\beq
\iint_{\R^2}f(\bq)g(\bp)\overline{\Pi}(\dd\bq,\dd\bp)
=f(\x)m(g)\hat{I}
\eeq
for any $f,g\in C(\R)$.
Then, if $g$ has a compact support we have $m(g)=0$.
It follows that $\overline{\Pi}(\R^2)=0$.
Thus, the fixed point is not a covariant POVM on $\R^{2}$.

\paragraph{Conclusion.}
This note has shown that BLW's proof \cite{BLW13} of Heisenberg's EDR fails,
since the proof wrongly concludes that for any measurement there is a covariant 
POVM obeying their EDR with marginals having at most the same error and 
disturbance.  This is not true for all the error-free linear measurements.
Thus, their claim is unsupported that Heisenberg's EDR can be 
proved in its original form with their new formulations of error and disturbance.

\acknowledgments
This work was supported by the MIC SCOPE, Grant Number 121806010,
by the John Templeton Foundations, Grant Number 35771,
and by the JSPS KAKENHI, Grant Numbers 21244007 and 24654021.

\end{document}